\documentclass[11pt,fleqn,twoside]{article}
\usepackage{amsfonts,amssymb,latexsym}
\makeatletter
\newcommand{\prava}{\footnotesize\it
\begin{flushright}
\begin{minipage}{18cm}
Copyright \copyright 1998 by Boris. A. Kupershmidt
\end{minipage}
\end{flushright}}

\newcommand{\name}[1]{\begin{flushleft}
                       \LARGE \bf #1
                       \end{flushleft}\vspace{-3mm}}

\newcommand{\Author}[1]{\begin{flushleft}
                       \it #1 \end{flushleft}}

\newcommand{\Adress}[1]{\begin{flushleft}
                       \it #1 \end{flushleft}}

\newcommand{\Date}[1]{\begin{flushleft}
                      \small  \it #1 \end{flushleft}}

\newcommand{\ehkol}{Author \ name}
\newcommand{\ohkol}{Article \ name}
\renewcommand{\@evenhead}{
\hspace*{-3pt}\raisebox{-15pt}[\headheight][0pt]{\vbox{\hbox to \textwidth
{\thepage \hfil \ehkol}\vskip4pt \hrule}}}
\renewcommand{\@oddhead}{
\hspace*{-3pt}\raisebox{-15pt}[\headheight][0pt]{\vbox{\hbox to \textwidth
{\ohkol \hfil \thepage}\vskip4pt\hrule}}}
\renewcommand{\@evenfoot}{}
\renewcommand{\@oddfoot}{}

\newcommand{\be}{\begin{equation}}
\newcommand{\ee}{\end{equation}}
\newcommand{\ba}{\hspace*{-5pt}\begin{array}}
\newcommand{\ea}{\end{array}}

\newcommand{\ds}{\displaystyle}
\makeatother

\begin{document}
\setcounter{page}{383}

\font\BoldMath=cmmib10 scaled \magstep1
\font\BoldMathN=cmbsy10 scaled \magstep1

\newcommand{\pmF}{\mbox{\BoldMath \char 70}}
\newcommand{\pmG}{\mbox{\BoldMath \char 71}}
\newcommand{\pmc}{\mbox{\BoldMath \char 99}}
\newcommand{\pmv}{\mbox{\BoldMath \char 118}}
\newcommand{\pmx}{\mbox{\BoldMath \char 120}}

\newcommand{\pmnabla}{\mbox{\BoldMathN \char 114}}

\thispagestyle{empty}

\renewcommand{\ehkol}{B.A. Kupershmidt}
\renewcommand{\ohkol}{Remarks on Random Evolutions in
Hamiltonian Representation}

\begin{flushleft}
\footnotesize \sf
Journal of Nonlinear Mathematical Physics \qquad 1998, V.5, N~4,
\pageref{kupershmidt_3-fp}--\pageref{kupershmidt_3-lp}.
\hfill {\sc Article}
\end{flushleft}

\vspace{-5mm}

\renewcommand{\footnoterule}{}
{\renewcommand{\thefootnote}{}
 \footnote{\prava}}

\name{Remarks on Random Evolutions \\ in Hamiltonian Representation}\label{kupershmidt_3-fp}
\Author{Boris A. KUPERSHMIDT}

\Adress{The University of Tennessee Space Institute, Tullahoma, TN
37388  USA\\
E-mail: bkupersh@utsi.edu}

\Date{Received April 27, 1998}

\begin{abstract}
\noindent
Abstract telegrapher's equations and some random walks of Poisson type are shown to f\/it into the framework of the Hamiltonian formalism after an appropriate time-dependent rescaling of the basic variables has been made. \end{abstract}

\setcounter{section}{1}
\setcounter{equation}{0}
\renewcommand{\theequation}{\arabic{section}.\arabic{equation}}

\section*{\S~1.  Introduction}
Time evolution of random processes dif\/fers in one essential respect from evolution of conservative systems in general and Hamiltonian systems in particular.  As a great number of various limit theorems attest, the f\/inal states loose all the information about the initial conditions.  Thus, there is nothing to be conserved, no constants of motion can exist, and no use can be made of the powerful machine of the Hamiltonian formalism.
Or so it seems.  Sometimes there {\it are} things which do not change
during time evolution, such as rates of decay and other universal
exponents.  This suggests that one may hope to f\/ind constants of
motion and even Hamiltonian forms in at least some probabilistic
systems provided one is willing to make
time-dependent rescalings of the basic dynamical variables.  Besides, there is something like a precendent in the history of attempts to quantize dissipative systems, a close relative of random processes.
The simplest of such systems is a particle moving on a line under the inf\/luence of a harmonic force and a friction:\be
 \ddot{x} + 2 k \dot{x} + bx = 0,
\ee where: $x=x(t)$, $x \in {\bf  R}^1$, is the position of the particle; $k$ and $b$ are constants; and overdot denotes the time-derivative.  Clearly, for $k \neq 0$ the equation (1.1) is not a Hamiltonian (or a Lagrangian) system {\it as it stands}.  However, set\be
x(t) = X(t)\;{\mathrm e}^{-kt}.
\ee Then\be
\ddot{x} + k \dot{x} + bx = \left[\ddot{X} + \left(b-k^2\right) X \right] {\mathrm e}^{-kt}
\ee and we get for $X$ the equation\be
\ddot{X} + \left(b - k^2\right) X = 0
\ee which is a Hamiltonian system with the Hamiltonian
\[
H (p,X) = {p^2 \over 2} + {\left(b-k^2\right) \over 2} X^2.
\] The main theme of this paper is that some (not all!) random
evolutions of Poisson type can be put into a Hamiltonian form after an
appropriate time-dependent rescaling of the basic variables has been
made.  This rescaling into a Hamiltonian form is, naturally, a
heuristic principle and not a general theorem.  We shall see  below
how this principle works and fails to work for a few representative
systems.  We start in the next Section with one-dimensional random
walk.  Multidimensional generalizations proceed
in two dif\/ferent directions:  abstract telegraphers's
equations, Section 3, or random walks in ${\bf  R}^d$ and ${\bf
Z}^d$,
Section 4.
\setcounter{section}{2}
\setcounter{equation}{0}
\renewcommand{\theequation}{\arabic{section}.\arabic{equation}}

\section*{\S~2. One-Dimensional Random Walk}

Let us consider a particle which moves on ${\bf R}^1$ with a constant speed $v$, and reverses direction according to a Poisson process with intensity $a$.  This model had been proposed by G.I.~Taylor~[1] in an attempt to understand turbulent dif\/fusion.  It is convenient to work with a discrete situation f\/irst, and then pass to the continuous limit; the reader will f\/ind a lucid analysis in Kac~[2] whose treatment I follow.
So, suppose we have a lattice ${\bf Z} \Delta x$.  Our particle moves with the speed $v$ in the positive or negative direction; after time $\Delta t = \Delta x/v$ it changes direction; with probability $1 - a \Delta t$ the direction stays the same.  Denote by $S_n$ the displacement of the particle after $n$ steps, with the initial step taken in the positive direction.  Given a function $\varphi = \varphi (x)$, def\/ine the expectation values
\setcounter{equation}{0}
\renewcommand{\theequation}{\arabic{section}.\arabic{equation}{\rm a}}\be
F_n^+ (x): = \langle \varphi (x + S_n) \rangle,
\ee  \setcounter{equation}{0}
\renewcommand{\theequation}{\arabic{section}.\arabic{equation}{\rm b}}
\be
F_n^- (x): = \langle \varphi (x - S_n) \rangle.\ee 
Thus, $F_n^+ (x)$ (resp. $F_n^- (x)$) is the expectation value of $\varphi (x)$ after $n$ steps, when the initial direction of the walk starting at $x$ is positive (resp. negative).  Considering the $n^{th}$ step as once removed from the $(n-1)^{st}$ one, we get in the usual way
\setcounter{equation}{1}
\renewcommand{\theequation}{\arabic{section}.\arabic{equation}}\be
F_n^\pm (x) = (1 - a \Delta t) F_{n-1}^\pm (x \pm v \Delta t) + a \Delta t F_{n-1}^\mp (x \pm v \Delta t)
\ee which can be suggestively rewritten as\be
\ba{l}
\ds {F_n^\pm (x) - F_{n-1}^\pm (x) \over \Delta t} = {F_{n-1}^\pm (x \pm v \Delta t) - F_{n-1}^\pm (x) \over \Delta t}  \\[4mm]\ds \qquad + a \left[F_{n-1}^\mp (x \pm v \Delta t) - F_{n-1}^\pm (x \pm v \Delta t) \right].
\ea
\ee Passing to the continuous limit, we obtain\be \left\{
\ba{l}
\ds {\partial F^+ \over \partial t} = v {\partial F^+ \over \partial x} - a (F^+ - F^-),\\[4mm]\ds {\partial F^- \over \partial t} = -v {\partial F^- \over \partial x} -a (F^- - F^+).
\ea \right.
\ee 
This is the dynamical system we were after.  The point of going through the discrete route f\/irst is the logical ease of deriving equation (2.2) (and similar equations later on).
Now comes the rescaling. Set\be
F^\pm = f^\pm {\mathrm e}^{-at}.
\ee Then the system (2.4) becomes\be
\left\{
\ba{l}
\ds {\partial f^+ \over \partial t}  = v {\partial f^+ \over \partial x} + af^-\\[4mm]
\ds {\partial f^- \over \partial t} = -v {\partial f^- \over \partial
x} + a f^+
\ea \right.
\ee and this is patently a Hamiltonian system since it can be written in the form\be
{\partial \over \partial t} \left(\begin{array}{c} f^+ \\[1mm]
f^-\end{array} \right)
= \left(\begin{array}{cc} v \partial & -a \\[1mm] a & v \partial\end{array}
\right) \left(\begin{array}{c} \delta H/\delta f^+ \\[1mm] \delta H/\delta
f^- \end{array} \right) \ee 
with\be
H = {(f^+)^2 - (f^-)^2 \over 2}
\ee and with\be
\partial: = \partial/\partial x.
\ee The matrix\be
\left( \begin{array}{cc}
v \partial  & - a \\[1mm] a  & v  \partial
\end{array}\right)
\ee is skewsymmetric constant-coef\/f\/icient and is, thus,
 Hamiltonian.  See,
e.g., [3], Ch.~I, for the modern point of view on Hamiltonian formalism; all the Hamiltonian matrices below are of this simple kind.
Note that the Hamiltonian $H$ (2.8) is the f\/irst in the inf\/inite series\be
H_n:  = {f^+ \partial^{2n} (f^+) - f^- \partial^{2n} (f^-) \over 2} , \qquad  n \in {\bf Z}_+
\ee of conserved densities of the system (2.6).  Indeed, writing\be
 h_1 \sim h_2
\ee when\be
(h_1 - h_2) \in Im \partial ,
\ee we have\[
\ba{l}
\ds  {\partial H_n \over \partial t} \sim {\delta H_n \over \delta f^+} \ {\partial f^+ \over \partial t} + {\delta H_n \over \delta f^-} \ {\partial f^- \over \partial t} \\[4mm]\ds \qquad  = \partial^{2n} (f^+) \left[v \partial (f^+) + a f^- \right] - \partial^{2n} (f^-) \left[-v \partial (f^-) + a f^+ \right] \\[4mm]\ds \qquad \sim \partial^{2n} (f^+) a f^{-} - \partial^{2n} (f^-) a f^+ \sim 0.
\ea
\]Thus, $H_n$ is a conserved density.  Moreover, it is obvious that all the $H_n$'s are in involution;\be
\{H_n, H_N \}: = X_{H_n} (H_N) \sim 0, \qquad \forall \; n, N \in {\bf Z}_+
\ee where $X_{H_n}$ is the evolution derivation corresponding to the f\/low with the Hamiltonian~$H_n$:\be
\ba{l}
\ds X_{H_n}
\left(\begin{array}{c}f^+ \\[1mm] f^-\end{array}
\right) = {\partial \over \partial t} \left(\begin{array}{c} f^+ \\ f^-\end{array}\right) =
\left(\begin{array}{cc}
v \partial  & - a \\[1mm] a  & v \partial
\end{array}
\right)
\ds \left(\begin{array}{c}\delta H_n / \delta f^+ \\[1mm]
\delta H_n / \delta f^-\end{array} \right) \\[6mm]\ds \qquad = \left(\begin{array}{c}
v \partial^{2n+1} (f^+) + a \partial^{2n} (f^-) \\[1mm]
 -v \partial^{2n +1} (f^-) + a \partial^{2n} (f^+)\end{array}\right).\ea
\ee 
Thus, we have an inf\/inite number of commuting f\/lows with an inf\/inity of commuting conserved densities.

We have considered the simplest possible system.  Before moving on to more general pastures, it is worthwhile to note that the same equations (2.2) arise for the pair of functions, $p^+ (x,t)$ and $p^- (x,t)$, describing the probability of f\/inding the particle at the point $x$ at the time $t$, arriving there from the right (for $p^+$) or left (for $p^-$) (see~[4], Ch.~I): \be
 p^\pm(x, t + \Delta t) = (1-a \Delta t) p^\pm (x \pm v \Delta t, t) + a \Delta t p^\mp (x \pm v t, t).
\ee In this form this equation is easy to generalize for the inhomogeneous case and even for the case when the particle is allowed to rest (see~[4], Ch.~I):\be
\ba{l}
p^\pm (x,t + \Delta t) \\[2mm]
\qquad = [1 - \sigma (x)] p^\pm (x,t)
+ [ \sigma (x \pm v \Delta t) - \lambda (x \pm v \Delta t) \Delta t] p^\pm (x \pm v \Delta t, t) \\[2mm]
\qquad +  \lambda (x \pm v \Delta t) \Delta t p^\mp (x \pm v \Delta t, t),
\ea
\ee where $\lambda(x)$ is the local intensity of the Poisson process, and $1- \sigma (x)$ is the local probability of resting.  Passing to the continuous limit we get\be
\left\{
\ba{l}
\ds {\partial p^+ \over \partial t}  = v {\partial \over \partial x} (\sigma (x) p^+) + \lambda (x) (p^- - p^+)\\[4mm]\ds {\partial p^- \over \partial t} = -v {\partial \over \partial x}
(\sigma (x) p^-) + \lambda (x) (p^+ - p^-).
\ea \right.
\ee If $\lambda$ (formerly $a$) is not a constant, we cannot renormalize the variables $p^\pm$ by ${\mathrm e}^{- \lambda t}$ since $t$ will enter {\it explicitly} into the motion equations; the system (2.18) in this case cannot be converted into a Hamiltonian form.  When, however, $\lambda$ is a constant, even though $\sigma$ (formerly 1) is not, a Hamiltonian form is possible. Set\be
 p^\pm ={\widetilde p}{\,}^\pm {\mathrm e}^{- \lambda t}.
\ee Then the system (2.18) becomes\be
\ba{l}
\ds {\partial \widetilde p^+ \over \partial t} = v {\partial \over
\partial x} (\sigma \widetilde p^+) + \lambda \widetilde p^- \\[4mm]\ds {\partial \widetilde p^- \over \partial t} = -v {\partial \over
\partial x}  (\sigma \widetilde p^-) + \lambda \widetilde p^+
\ea
\ee which can be rewritten as\be
{\partial \over \partial t}
\left( \begin{array}{c}
\widetilde p^+ \\[1mm] \widetilde p^- \end{array} \right) = \left(\begin{array}{cc}
v \partial  & - \lambda \sigma^{-1} \\[1mm]
\lambda \sigma^{-1}  & v \partial \end{array} \right)
\left( \begin{array}{c} \delta H / \delta \widetilde p^+ \\[1mm] \delta H / \delta \widetilde p^-\end{array} \right)
\ee with\be
H = \sigma  {(\widetilde p^+)^2 - (\widetilde p^-)^2 \over 2}.
\ee 
\setcounter{section}{3}
\setcounter{equation}{0}
\renewcommand{\theequation}{\arabic{section}.\arabic{equation}}

\section*{\S~3. Telegrapher's Equation}
The system (2.4) is 2-component f\/irst-order in time. ``Now the amazing thing is that these two linear equations of f\/irst order can be combined into a [single] hyperbolic equation", says Kac ([2], p.~500), and proceeds as follows.  Set\be
F: = {1 \over 2} (F^+ + F^-), \qquad G: = {1 \over 2} (F^+ - F^-) \ee so that
\setcounter{equation}{1}
\renewcommand{\theequation}{\arabic{section}.\arabic{equation}{\rm a}}\be
\ds {\partial F  \over  \partial t}= v {\partial G  \over \partial x}
\ee 
\setcounter{equation}{1}
\renewcommand{\theequation}{\arabic{section}.\arabic{equation}{\rm b}}\be
{\partial G \over  \partial t}= v {\partial F  \over  \partial x} -2a G
\ee whence\setcounter{equation}{2}
\renewcommand{\theequation}{\arabic{section}.\arabic{equation}}
\be
{\partial^2F \over \partial t^2} + 2a {\partial F \over \partial t} = v^2 {\partial^2 F \over \partial x^2}
\ee which is the telegrapher's equation.  Rewritten as\be
v^{-2}  {\partial^2 F \over \partial t^2}  +  {2a \over v^2}  {\partial F \over \partial t}  =  {\partial^2 F \over \partial x^2}
\ee it can be considered as a singular perturbation of the dif\/fusion equation\be
{1 \over D}{\partial F \over \partial t} = {\partial^2 F \over \partial x^2}
\ee where\[
{1 \over D} = \lim  {2a \over v^2}
\]when both $a$ and $v$ tend to inf\/inity.  Now, the dif\/fusion equation assumes unlimited speeds of microscopic agents, clearly an untenable thesis in view of special relativity.  The hyperbolic equation (3.4)can be considered then as a sort of relativistic regularization of the classical dif\/fusion and heat equations.
Let us now look at Hamiltonian properties of this equation, but f\/irst we generalize it to the form\be
\epsilon {\partial^2 u \over \partial t^2} + 2a {\partial u \over \partial t} = L(u)
\ee where $L$ is an arbitrary linear selfadjoint operator in arbitrary
number of  space dimensions:\be
L^\dagger = L
\ee and $\epsilon$ is a constant (considered small if desired).
The case\be
L = A^2
\ee where $A$ is a skewadjoint operator:\be
 A^\dagger = -A
\ee is the most direct generalization of dif\/ferential equations of telegrapher's type to which probabilistic interpretation applies~[5]; more about this case later on.
Set\be
u = \overline{u}\; {\mathrm e}^{\lambda t}
\ee where $\lambda$ is a constant to be specif\/ied presently.  Since
\[
 {\mathrm e}^{- \lambda t} \left(\epsilon  {\partial^2 \over \partial
t^2} +
2a {\partial \over \partial t}\right) {\mathrm e}^{\lambda t}
= \epsilon {\partial^2 \over \partial t^2} + 2 (\lambda \epsilon + a)
{\partial \over \partial t} + (\epsilon \lambda^2 + 2a \lambda),
\]
choosing\be
\lambda = -a/\epsilon
\ee we transform equation (3.6) into equation\be
\epsilon {\partial^2 \overline{u} \over \partial t^2} =
\widehat{L} (\overline{u}),
\ee  where\be
 \widehat{L}: = L + {a^2 \over \epsilon}
\ee is again a selfadjoint operator.  The second-order equation (3.12), written as a f\/irst-order system\be
\left\{
\ba{l}
\ds {\partial \overline{u} \over \partial t}  = \widetilde{u}\\[3mm]\ds {\partial \widetilde{u} \over \partial t}  = \epsilon^{-1} \widehat{L} (\overline{u})
\ea \right.
\ee is easily seen to be a canonical Hamiltonian system:\be
{\partial \over \partial t}
\left( \begin{array}{c}\overline{u} \\[1mm]\widetilde{u}\end{array}\right)  = \left(
\begin{array}{cc}0  & 1 \\[1mm] -1  & 0\end{array}\right) \left( \begin{array}{c}
\delta H / \delta \overline{u} \\[1mm]
 \delta H / \delta \widetilde{u}\end{array}\right)
\ee with\be
 H = H_0 = {\widetilde{u}^2 \over 2} - {1 \over 2 \epsilon}  \overline{u} \widehat{L} (\overline{u}).
\ee (It is in this place that the selfadjointness of $\widehat{L}$ plays a r$\hat{{\rm o}}$le).  Like for the system (2.6), we have an inf\/inity of commuting conserved densities for the system (3.14):\be
H_n = {1 \over 2} \widetilde{u} \widehat{L}^n (\widetilde{u}) - {1 \over 2 \epsilon} \overline{u} \widehat{L}^{n + 1} ( \overline{u}), \qquad n \in {\bf Z}_+.
\ee 
The alert reader may have noticed that the Hamiltonian form (2.7) of  1-dimensional random walk (2.4) is {\it different} from the canonical Hamiltonian form (3.15) of its generalization (3.6).  How could this  have happenend?  The ultimate reason is that the {\it system}~(2.4) is more rigid than the {\it
scalar} second-order equation (3.6):  the latter can be written in a multitude of ways as a 2-component f\/irst order system.  For example, a direct generalization of the Hamiltonian form (2.7) exists for the case when $L=A^2$ with a skewadjoint~$A$.  Then equation (3.12) results from the following Hamiltonian system:
\be
\left(
\begin{array}{c}
\ds {\partial \overline{u} \strut \over \strut \partial t} \\[4mm]\ds {\partial \widetilde{u} \strut \over \strut \partial t}
\end{array} \right)= \left(
\begin{array}{c}
\ds X(\overline{u}) + {a \over \epsilon} \widetilde{u} \\[3mm]
\ds  -X(\widetilde{u})  + {a \over \epsilon} \overline{u}\end{array}
\right) =
\left( \begin{array}{cc}X  & -a  \epsilon^{-1} \\[2mm]a \epsilon^{-1}  & X\end{array}\right)
\left( \begin{array}{c}
\overline{u} \\[1mm]
 - \widetilde{u}\end{array}
\right)\ee 
\be
\qquad  = \left(
\begin{array}{cc}
X &  -a \epsilon^{-1} \\[2mm]
 a \epsilon^{-1}  & X\end{array} \right)
\left(\begin{array}{c} \delta / \delta \overline{u} \\[1mm]
 \delta / \delta \widetilde{u}\end{array}\right)
 \left({\overline{u}^2 - \widetilde{u}^2 \over 2}\right)
\ee where\be
X = \pm A \epsilon^{-1/2}.
\ee Again,\be
 H_n = {1 \over 2} \overline{u} X^{2n} ( \overline{u}) - {1 \over 2} \widetilde{u} X^{2n} ( \widetilde{u}), \qquad  n \in {\bf Z}_+, \ee 
is an inf\/inite commuting set of conserved densities of the system (3.18).  In addition, the same equation (3.12) results from the following Hamiltonian system, quite {\it different} from~(3.18):\be
\left( \begin{array}{c} \ds
{\partial \overline{u} \strut \over \strut \partial t} \\[4mm]\ds {\partial \widetilde{u} \strut \over \strut \partial t}
\end{array} \right) = \left(\begin{array}{c}\ds
X(\widetilde{u}) + {a \over \epsilon} \overline{u} \\[3mm]
\ds  X (\overline{u})  - {a \over \epsilon} \widetilde{u}
\end{array}\right) = \left( \begin{array}{cc}X & a \epsilon^{-1}
\\[2mm] -a \epsilon^{-1}  & X \end{array}\right)
\left( \begin{array}{c} \widetilde{u} \\[1mm]
 \overline{u}\end{array}\right) \ee \be
\qquad  = \left( \begin{array}{cc} X  & a \epsilon^{-1} \\[2mm]
 -a \epsilon^{-1}  & X \end{array}\right)
\left( \begin{array}{c}
\delta / \delta \overline{u} \\[2mm]
\delta / \delta \widetilde{u}
\end{array}\right) \left( \overline{u}\widetilde{u} \right).
\ee In this case, an inf\/inity of commuting conserved densities is given by the formula\be
H_n = \overline{u} X^{2n} (\widetilde{u}), \qquad n \in {\bf Z}_+.
\ee 

\setcounter{section}{4}
\setcounter{equation}{0}
\renewcommand{\theequation}{\arabic{section}.\arabic{equation}}

\section*{\S~4. Multidimensional Random Walk}
A particle moves in ${\bf R}^d$ with a constant velocity
$\pmv \in \{\pmv_i \}$.  After each time interval $\Delta t$, there
is a change of velocity.  The change from $\pmv_i$ to $\pmv_j$ has
the probability $p_{ij}$, and we take\be
p_{ij} = \delta_{ij} + \alpha_{ij} \Delta t, \ee 
with\be
\sum_j \alpha_{ij} = 0, \qquad \forall \; i.
\ee (In the continuous limit, for the set of states $\{\pmv_i\}$ we have a Markov process $\xi (t)$ with the transition probabilities\be
p_{ij} (\Delta t) = \delta_{ij} + \alpha_{ij} \Delta t + {\rm o} (\Delta t),
\ee but, as in  \S~1, it is more convenient to start with the discrete picture.)  Denoting by $F_i = F_i (\pmx, t)$ the probability of f\/inding the particle coming for its velocity change into the point $\pmx$ at time $t$ with the velocity $\pmv_i$, we have, similar to \S~1,\be
 F_i (\pmx, t + \Delta t) = \sum_j p_{ji} F_j (\pmx - \pmv_i \Delta t, t).
\ee By virtue of formula (4.1), in the continuous limit we get \be
{\partial F_i \over \partial t} = - (\pmv_i \cdot \pmnabla) (F_i) + \sum_j \alpha_{ji} F_j
\ee 
where\be
\pmv_i \cdot \pmnabla: = \sum_{s=1}^d (v_i)_s {\partial \over \partial x_s}.
\ee Set\be
F_i = f_i \;{\mathrm e}^{- \lambda t}.
\ee Then equation (4.5) becomes\be
{\partial f_i \over \partial t} = - (\pmv_i \cdot \pmnabla) (f_i) + \sum_j \beta_{ji} f_{j}
\ee 
where\be
\beta_{ji}: = \alpha_{ji} + \lambda  \delta_{ij}
\ee so that the constraint (4.2) turns into\be
\sum_j \beta_{ij} = \lambda, \qquad \forall \; i .
\ee 
We are going to analyze the system (4.8), (4.10) from the Hamiltonian point of view.  As the Hamiltonian we pick\be
H = {1 \over 2} \sum_i c_i (f_i)^2
\ee 
with some unknown constants $c_i$'s.  The constant-coef\/f\/icient Hamiltonian matrix\be
B_{ij} = - \delta_{ij} {1 \over c_i} \pmv_i \cdot \pmnabla + \Gamma_{ij}
\ee where $\Gamma = (\Gamma_{ij})$ is a constant skewsymmetric matrix, reproduces the motion equa\-tions (4.8) through the Hamiltonian ansatz\be
{\partial f_i \over \partial t} = \sum_j B_{ij} \left({\delta H \over \delta f_j}\right)
\ee if\/f\be
 \beta_{ji} = \Gamma_{ij} c_j \qquad ({\rm{no \ sum \ on}} \ j).
\ee 
Let us estimate the proportion of Hamiltonian random walks among all of them.  The dimension of the latter is the dimension of the space of the $\beta$'s subject to the conditions~(4.10), but with the understanding that $\lambda$ is at our disposal.  Thus,\be
{\rm{Total \ dim}} \; = N^2 - N + 1
\ee where $N$ is the number of the dif\/ferent $f_i$'s (and also the
number of the velocities $\pmv_i$'s).  From (4.10) and (4.14)
we get
\[
 \lambda = \sum_i \beta_{ji}
= \sum_i \Gamma_{ij} c_j = \left(\sum_i \Gamma_{ij}\right) c_j 
\]
so that\be
 c_j = {\lambda \over \sum_i \Gamma_{ij}} \ee 
(or no conditions for $\lambda = 0$).  Thus, we have to look at the dimension of the image of the map $\Gamma \times \pmc \longmapsto (\Gamma \widehat{c})^t = \beta$, where\be
\pmc: = (c_1, \ldots , c_N)^t, \qquad \widehat{c}: = \; {\rm{diag}} \; (c_1, \ldots , c_N),
\ee and where $\pmc$ is a function of $\Gamma$ given by formula (4.16).  Let us compute this dimension at the point in the $\beta$-space which corresponds to\be
\Gamma_0 = \left( \begin{array}{cc}
{\bf 0}  & {\bf 1} \\[1mm]
 -{\bf 1}  & {\bf 0}\end{array}\right);
\ee thus, we assume that $N$ is even:
\be
 N = 2 \overline{N}.
\ee 
Then\be
\widehat{c}_0 = \lambda \left(
\begin{array}{cc}
- {\bf 1}  & {\bf 0} \\[1mm]{\bf 0}  & {\bf 1}\end{array}\right)
\ee and\be
 \beta_0^t = \lambda \left(
\begin{array}{cc}
 {\bf 0}  & {\bf 1} \\[1mm]{\bf 1} &  {\bf 0}\end{array}\right).
\ee (This $\beta$ corresponds to a direct sum of one-dimensional random walks.)
Let\be
\Gamma = \Gamma_0 + \epsilon \overline{\Gamma}
\ee \be
c = c_0 + \epsilon \overline{c}
\ee \be
 \beta^t = \beta_0^t + \epsilon \overline{\beta}^t \ee 
be an inf\/initesimal change of the objects under consideration (i.e., $\epsilon^2 = 0$).  Denote\be
\gamma_j: = \sum_i \overline{\Gamma}_{ij}
\ee \be
\overline{\Gamma} = \left( \begin{array}{cc}
{\cal A}  & {\cal B} \\[1mm]-{\cal B}^t  & {\cal C} \end{array} \right), \qquad
{\cal A}^t = -{\cal A}, \quad {\cal C}^t = -{\cal C}.
\ee Then\be
\widehat{c} = \lambda
\left( \begin{array}{cc}
- {\bf 1}  & {\bf 0} \\[1mm] {\bf 0}  & {\bf 1} \end{array}\right)
 - \lambda \epsilon \; {\rm{diag}} \;(\gamma_1, \ldots, \gamma_N)
\ee and hence\be
\lambda^{-1} \overline{\beta}^t = \pmatrix{&&&- \gamma_{\overline{N} +1} & & \cr&&&&\ddots&\cr &&&&& - \gamma_{2 \overline{N}}\cr\gamma_1&&&&&\cr&\ddots&&&&\cr&&\gamma_{\overline{N}}&&&\cr} +
\left( \begin{array}{cc}
-{\cal A}  & {\cal B} \\[1mm]
 {\cal B}^t  & {\cal C} \end{array} \right).
\ee Taking ${\cal A}$ and ${\cal C}$ out of
$\lambda^{-1} \overline{\beta}^t$, we are left with the matrix\be
{\cal B} - \; {\rm{diag}} \; \left(\sum_i {\cal B}_{i 1} , \ldots , \sum_i {\cal B}_{i \overline{N}}\right)
\ee which amounts to an arbitrary matrix $\widehat{\cal B}$ subject to the conditions\be
\sum_i \widehat{\cal B}_{ij} = 0, \qquad \forall \; j .
\ee Thus, the dimension of the $\overline{\beta}$'s is\be
 {\overline{N}^2 - \overline{N} \over 2}  +  {\overline{N}^2 - N \over 2}  +  (\overline{N}^2 - \overline{N}) = 2 (\overline{N}^2 - \overline{N}) = {N^2 \over 2} - N. \ee 

Taking into account our free  parameter $\lambda$, we f\/inally get the dimension of the space of Hamiltonian random walks around the point $\beta_0$:\be
{\rm{Ham \ dim}} \; = {N^2 \over 2} - N + 1
\ee which is more than half the total dimension (4.15) of the space of random walks.
I conclude with a few remarks.
\medskip

\setcounter{equation}{33}

\noindent
{\bf Remark 4.33.}  Even for $d=1$ the random walk model in this section is more general than the one-dimensional model considered in \S~1, since we allow $N (=2 \overline{N})$ dif\/ferent velocities rather than two.  The case $d=1$ is special, having all possible velocities being proportional to each other.  This fact leads to an existence of an inf\/inity of commuting conserved densities for the system (4.8):\be
H_n = {1 \over 2} \sum_i c_i [\partial^n (f_i)]^2. \ee 

\setcounter{equation}{35}

\noindent
{\bf Remark 4.35.}  In our calculations of dimensions we had no use for the convective terms $(\pmv_i \cdot \pmnabla) (f_i)$.  Had these terms been absent to begin with, e.g., for $\pmx$-independent solutions, we would have been dealing with a system of ordinary dif\/ferential equations for the $f_i$'s; formula (4.32) in this case  provides a (very) low bound on the dimension of such systems with a particular Hamiltonian representation.  Such systems\be
 \dot{\pmF}{\,}^t = \pmF{\,}^t \beta
\ee where\[
\pmF{\,}^t: = (F_1, \ldots , F_N), \qquad \beta: = (\beta_{ij})
\]appear, e.g., for column-sums of the inverse Kolmogorov equation for a Markov process with transition probabilities (4.3):\be
 {d \over dt} \ p_{ij} = \sum_k p_{ik} \beta_{kj}
\ee where\be
\beta_{kj}:= \alpha_{kj} + \lambda_j \delta_{kj} \ee 
and\be
 F_j:= \sum_i p_{ij}.
\ee Hamiltonian analysis can be applied to the full system (4.37) and also to the direct Kolmogorov equation\be
 {d \over dt} \ p_{ij} = \sum_k \beta_{ik} p_{kj}.
\ee 

\setcounter{equation}{41}

\noindent
{\bf Remark 4.41.}  If the set of all possible velocities $\{\pmv_i\}$ is not discrete but is a continuous one, the sum sign in equation (4.8) turns into an integral sign.  The Hamiltonian arguments undergo a similar minor modif\/ication.
\medskip

\setcounter{equation}{42}

\noindent
{\bf Remark 4.42.}  If the randomly walking particle has internal
degrees of
freedom [6], equation (4.4) changes into\be
 F_i^\mu (\pmx, t + \Delta t) = \sum p_{ji}^{\nu \mu} F_j^\nu (\pmx - \pmv_i \Delta t, t)
\ee with\be
p_{ij}^{\mu \nu} = \delta_{ij} \delta_{\mu \nu} + \alpha_{ij}^{\mu \nu} \Delta t
\ee where indices $\mu, \nu$ refer to the internal states.  Equation (4.5) then becomes\be
 {\partial F_i^\mu \over \partial t} = - (\pmv_i \cdot \pmnabla) (F_i^\mu) + \sum_{j, \nu} \alpha_{ji}^{\nu \mu} F_j^\nu
\ee with\be
\sum_{i, \mu} \alpha_{ji}^{\nu \mu} = 0, \qquad\forall \; j, \nu.
\ee The Hamiltonian analysis proceeds as before, with the quadratic Hamiltonian now being\be
 H = {1 \over 2} \sum_{i_, \mu_, \nu} c_i^{\mu \nu} F_i^\mu F_i^\nu.
\ee In the simplest one-dimensional case, where $\{\pmv_i \} = \{\pm \pmv \}$, and where everything is invariant with respect to the ref\/lection $x \mapsto - x$, as in \S~1, we have\be
\left\{
\ba{l}
\ds {\partial \pmF_+ \over \partial t}  = v {\partial \pmF_+ \over \partial x} + \widetilde{\alpha} \pmF_+ + \overline{\alpha} \pmF_- \\[4mm]\ds {\partial \pmF_- \over \partial t}  = -v {\partial \pmF_- \over \partial x} + \widetilde{\alpha} \pmF_- + \overline{\alpha} \pmF_+
\ea \right.
\ee where\[
(\pmF_\pm)^\mu = (F_\pm^\mu), \qquad \widetilde{\alpha}^{\mu \nu}:= \alpha_{+ +}^{\nu \mu} = \alpha_{- -}^{\nu \mu}, \qquad\overline{\alpha}^{\mu \nu}: = \alpha_{+ -}^{\nu \mu} = \alpha_{- +}^{\nu \mu},
\]etc.  We also have a vector analog of the telegrapher's equation:  Set\be
\pmF:= {1 \over 2} (\pmF_+ + \pmF_-), \qquad\pmG: = {1 \over 2} (\pmF_+ - \pmF_-)
\ee so that\be
\left\{
\ba{l}
\ds {\partial \pmF \over \partial t} = v {\partial \pmG \over \partial x} + (\widetilde{\alpha} + \overline{\alpha}) \pmF \\[4mm]\ds {\partial \pmG \over \partial t} = v {\partial \pmF \over \partial x} + (\widetilde{\alpha} - \overline{\alpha})\pmG.
\ea \right.
\ee 
Then\be
 {\partial^2 \pmF \over \partial t^2} - 2 \widetilde{\alpha}  {\partial \pmF \over \partial t} = v^2  {\partial^2 \pmF \over \partial x^2} + \left(\overline{\alpha}^2 - \widetilde{\alpha}^2 + [\overline{\alpha}, \widetilde{\alpha}]\right) \pmF
\ee where the matrices $\widetilde{\alpha}$ and $\overline{\alpha}$ are subject to the condition\be
(1, \ldots , 1) (\widetilde{\alpha} + \overline{\alpha}) = (0, \ldots , 0).
\ee 

\label{kupershmidt_3-lp}

\end{document}